# Resonance spin-charge phenomena and mechanism of magnetoresistance anisotropy in manganite/metal bilayer structures

V.A. Atsarkin, B.V. Sorokin, I.V. Borisenko, V.V. Demidov, and G.A. Ovsyannikov

Kotel'nikov Institute of Radio Engineering and Electronics of RAS

11-7, Mokhovaya Str., 125009 Moscow, Russia

Abstract

The dc voltage generated under ferromagnetic resonance has been studied in bilayer structures based on manganite thin epitaxial films La$_{0.67}$Sr$_{0.33}$MnO$_3$ (LSMO) and non-magnetic metals (Au, Pt, and SrRuO$_3$) in the temperature range up to the Curie point. The effect is shown to be caused by two different phenomena: (1) the resonance dc electromotive force related to anisotropic magnetoresistance (AMR) in the manganite film and (2) pure spin current (spin pumping) registered by means of the inverse spin Hall effect in normal metal. The two phenomena were separated using the angular dependence of the effect, the external magnetic field H$_0$ being rotated in the film plane. It was found that the AMR mechanism in the manganite films differs substantially from that in traditional ferromagnetic metals being governed by the colossal magnetoresistance together with the in-plane magnetic anisotropy. The spin pumping effect registered in the bilayers was found to be much lower than that reported for common ferromagnets; possible reasons are discussed.

1. Introduction

Excitation of ferromagnetic resonance (FMR) in ferromagnetic metals (FM) results in several interesting spin-charge effects, such as a change in electrical resistance [1-6], appearance of dc electromotive force (the resonance e.m.f. effect) [1, 2, 7-10], and flowing of pure spin current across the interface of FM with adjacent non-magnetic ("normal") metal (NM) [9, 11-16]. Recently, the resonance spin-charge phenomena attract heightened attention due to both pure physical interest and prospects of application in spintronics [17, 18]. As a rule, the studies in this field have been performed with common magnetic metals and alloys; the results were explained successively in terms of standard ideas on the anisotropic magnetoresistance (AMR) and various manifestations of the spin Hall effect.

Much less is known, however, on the resonance spin-charge effects in the doped rare-earth manganites possessing unique magnetic and transport properties prospective for many applications (see, for example, the review articles [19-21] and references therein). The physical picture of the spin-charge interplay in manganites is strongly complicated due to such features as colossal magnetoresistance (CMR), metal-insulator transition near the Curie point ($T_C$), rich phase diagram, tendency to phase segregation, etc. Particularly, it was ascertained recently that the AMR effect in the rare-earth manganites differs considerably from that in common FMs in its magnitude as well as temperature and angular dependences. Apparently, this points to some peculiarities in the physical mechanism of the effect. The theory was proposed [22] based on quantum-mechanical calculation of the spin-orbit interaction with account made for the crystalline field; as a result, some experimental data were explained. However, this approach is limited to the case of low temperatures ($T \ll T_C$) and still was not supported sufficiently in a wide range of temperatures and manganite compositions.

The so-called pure spin current is one of the most interesting and prospective effects arising under conditions of the FMR excitation [11, 12]. The pure spin current is not related with transfer of electric charge and corresponds to the flow of the non-equilibrium component of the spin momentum from the excited ferromagnet to adjacent NM. FI/NM structures were also approved for the spin current generation (here FI stands for ferromagnetic insulator) [23, 24, 25]. It was shown [11, 12, 17] that the spin-dependent value of the spin current $\mathbf{I}_s$ is proportional to

[**m**×d**m**/dt], where **m** is the unit vector directed along the ferromagnetic moment **M** precessing at the FMR frequency. The cited expression is analogous to the Gilbert relaxation term in the Landau-Lifshits equation. As a consequence, spin current affects the relaxation width of the FMR line which increases markedly due to the flowing out of the transverse magnetization component across the interface with NM. This effect can be used for estimation of $\mathbf{I}_s$ [11]. The most reliable method of detecting and measuring the pure spin current is based on registering the dc voltage $U^{SP}$ (the superscript stands for spin pumping, SP) provided by the inverse spin Hall effect (ISHE) in the NM accepting the spin flow [13]. The magnitude of $U^{SP}$ depends on the spin current generation rate, the interface transparency for various spin components (the spin conductance tensor), and the ISHE value in the NM. The latter is determined mostly by the spin-orbit interaction efficiency which increases steeply in heavy metals. Up to date, the maximum values of $U^{SP}$ (up to several mV) were obtained in the FI/NM bilayers where the $Y_3Fe_5O_{12}$ (YIG) epitaxial film was used as the source of the spin current and thin films of Pt, Ta, W, etc. served for the ISHE registering [24, 25].

In the FM/NM structures employing standard ferromagnetic metals such as Fe, Py, etc., the $U^{SP}$ magnitude was found to be considerably less [9, 15, 16]. Besides, in this case one has to separate the measured dc voltage into two components: $U^{AMR}$ and $U^{SP}$ arising in FM and NM, respectively. The reliable method of distinguishing these contributions is described in Ref. [9]; it employs the specific dependence of $U^{AMR}$ on the angle between the external magnetic field $\mathbf{H_0}$ and the direction of the microwave current induced in the FM film by the resonance pumping. This technique is widely used; particularly, it was employed in our previous work [10] concerned with the resonance e.m.f. effect in the manganite thin films. In this case, however, a problem arises because of lack of information about the AMR mechanism in manganites, see above. Thus, it seems to be reasonable to continue and advance studies of the resonance spin-charge phenomena in thin manganite films and bilayer structures, including the detection of pure spin current and new analysis of the AMR mechanism accounting for the CMR effect present in manganites. These issues are the subject of the present work.

## 2. Experimental

The samples under study were bilayer structures LSMO/Au, LSMO/Pt and LSMO/SrRuO$_3$, based on thin epitaxial films of $La_{0.67}Sr_{0.33}MnO_3$ (LSMO). The SrRuO$_3$ (SRO) metal is paramagnetic above 160 K; it has good conductivity and allows easy epitaxial growth on perovskite substrates [26]. The LSMO films with the thickness of 50 – 100 nm were epitaxially grown on the (011) surface of the 5×5×0.5 mm NdGaO$_3$ (NGO) substrate by laser ablation; for detailed description see Refs. [27, 28]. To obtain the FM/NM bilayers, Au was deposited over the LSMO film *in situ*, whereas Pt was deposited over LSMO *ex situ*. The LSMO/SRO bilayer films were epitaxially grown on the same substrate by RF magnetron sputtering [29]. It is known [27, 30] that the basic plane of the LSMO films grown on the orthorhombic (110)NGO substrate is (001) (we use the pseudocubic notation for LSMO), and an additional in-plane uniaxial magnetic anisotropy is created with the easy axis directed along $[010]_{LSMO}$ due to orthorhombic distortions of the LSMO crystal structure. In the samples under study, the substrate plates were cut by such a way that the LSMO easy axis (corresponding to the $[1\bar{1}0]$ direction of the NGO substrate) coincided with an edge of the substrate for LSMO/Au and LSMO/Pt bilayers and was declined from the edge by 40 deg. for LSMO/SRO one (Fig. 1).

The experiments were carried out in the temperature range of 295-360 K. The temperature was measured by use of preliminary calibrated film resistance R(T). Thus, the film under study served as a self-thermometer, allowing to take into account additional heating produced by measuring current and microwave pumping. Typical R(T) data for similar LSMO films are reported in Ref. [6].

The main set of experiments was performed with a home-made EPR spectrometer which allowed electrical voltage measuring and additional irradiation of the sample with high power microwave radiation. The sample was situated horizontally into the central maximum of the microwave magnetic field **h** of the rectangular $TE_{102}$ cavity with the loaded quality factor $Q \sim 10^3$ at the frequency $\omega/2\pi \sim 9$ GHz. The external magnetic field $\mathbf{H}_0$ could be rotated around the vertical axis, thus remaining in the film plane. Contact platinum electrodes were sputtered at four corners or along opposite sides of the manganite film or covering metal, see Fig. 1.

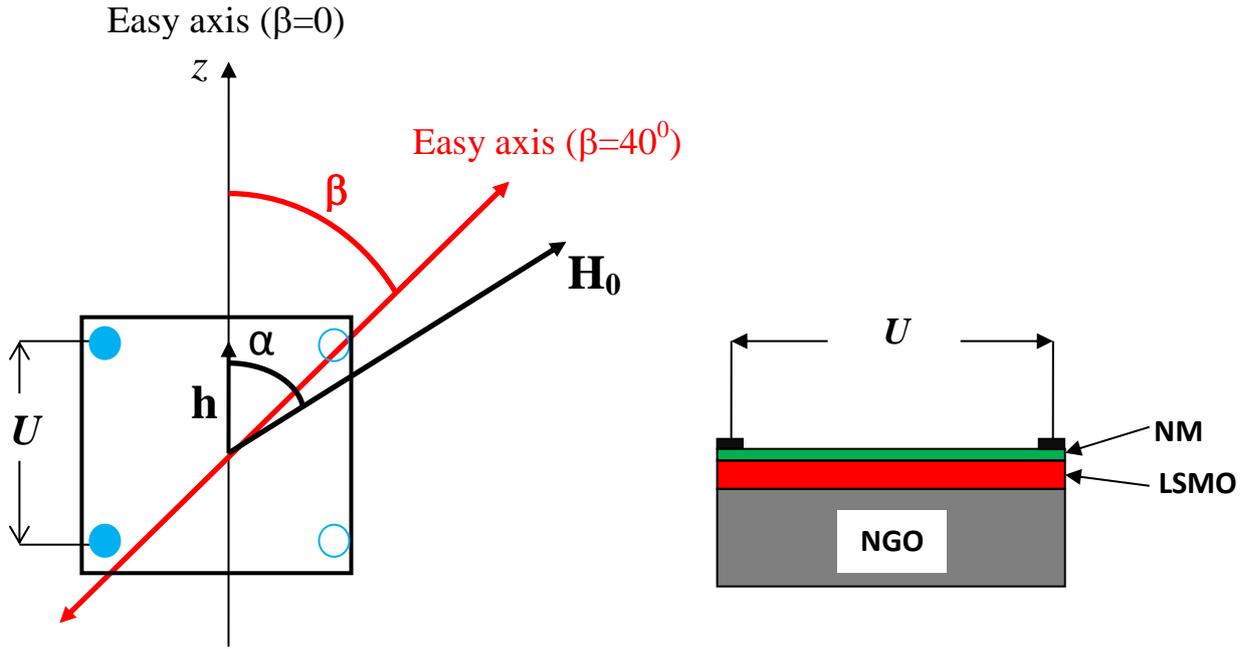

**Fig. 1.** The sketch of experimental geometry. The easy axis directions are shown at the left panel for two types of samples (see the text).

As a rule, the voltage U was measured along the sample border parallel to the wider wall of the cavity, that is along **h** (z axis in Fig.1). Nominally, microwave currents cannot be induced in this direction. In practice, however, the $TE_{102}$ mode is perturbed by the sample and leads, so the z-component of microwave current does exist and provides considerable dc effect under study [10].

The microwave pumping with the power P up to 250 mW was supplied by the Gann diodes. To distinguish the effect of resonant pumping, the microwave power was square-wave modulated at the modulation frequency $f_m$ = 100 kHz. Correspondingly, the voltage U was lock-in amplified and detected with the reference frequency $f_m$. To enhance sensitivity, accumulation was used at repeated **H₀** sweeping across the FMR line. With this technique, the signals down to 0.01 μV were reliably recorded. The U values given below are converted to dc voltage at the contact electrodes.

The external magnet was supplied through a switch key allowing inversion of the field direction. Since the dc signals caused both by AMR and spin pumping mechanisms change the sign together with the change of the **H₀** polarity, hereafter the half-difference of the signals measured at opposite field directions are presented. This trick enables one to avoid any parasitic signals which do not depend on the field direction [10].

The same setup could be employed for FMR spectroscopy. When necessary, more precise spectral measurements were performed at the commercial X-range EPR spectrometer Bruker ER 200. The angular dependences of the FMR spectra were used for determination of the main parameters of the LSMO layers, such as saturated magnetization M, uniaxial magnetic anisotropy field $H_u$ and the FMR line width $\Delta H$ (the absorption full width at half maximum), see Table 1. Generally, the $H_u$ values at room temperature were about hundreds Oe, exceeding strongly the natural crystallographic (cubic) anisotropy of LSMO, so the latter was disregarded. Note that the FMR line-width may serve as a sensitive indicator of the sample quality, including the film homogeneity and possible crystal twinning. In the best LSMO films, the value of $\Delta H$ did not exceed 20 Oe at 295 K, and passed through its maximum of about 200 Oe near the Curie point ($T_C$ = 330 -350 K).

**Table 1.** Main parameters and experimental data obtained with the bilayer samples under study. $U^{AMR}_{max}$ and $U^{SP}_{max}$ are the peak absolute values of dc signals caused by the anisotropic magnetoresistance and spin pumping effects, respectively. The data shown in six right columns correspond to the room temperature.

| Structure | $t_{FM}/t_{NM}$ (nm/nm) | $T_C$ (K) | M (Oe) | $H_u$ (Oe) | $\Delta H$ (Oe) | $\beta$ (deg.) | $U^{AMR}_{max}$ ($\mu V$) | $U^{SP}_{max}$ ($\mu V$) |
|---|---|---|---|---|---|---|---|---|
| LSMO/Au | 65/15 | 347 | 285 | 85 | 28 | 0 | 0.13 | 0.34 |
| LSMO/Pt | 80/10 | 347 | 302 | 139 | 20 | 0 | 0.2 | 0.5 |
| LSMO/SRO | 60/10 | 331 | 271 | 101 | 80 | 40 | 1.2 | 0.55 |

3. Results

Typical signals $U(H_0)$ recorded in the samples LSMO/Au and LSMO/SRO by passing the magnetic field across FMR conditions are shown in Fig. 2 (insets). At fixed temperature, the signal magnitudes were found to be proportional to the microwave power P within the available range.

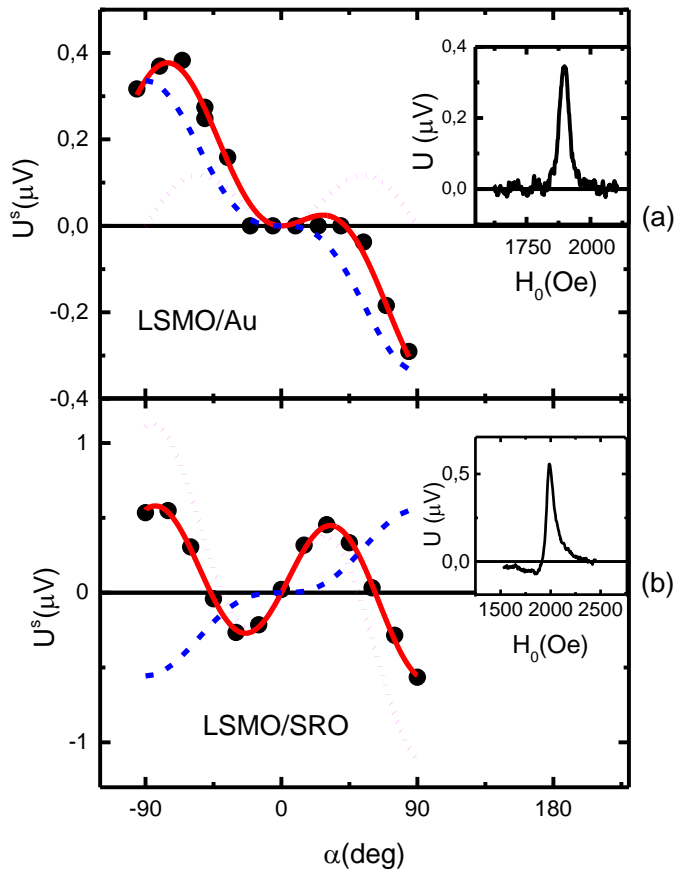

**Fig. 2.** Magnitude of symmetric part $U^s$ of the resonance dc voltage as a function of the angle $\alpha$ between the field direction and the sample edge for (a) LSMO/Au and (b) LSMO/SRO. Filled circles are experimental data. Solid curves are the best fits to Eq. (1) with $\beta=0$ (a) and $\beta=40^0$ (b). Dotted and dashed curves correspond to the AMR and SP effects, respectively. Insets: typical $U(H_0)$ signals recorded upon FMR pumping.

It is expected that these signals are resulted from two different effects: (1) rectification of microwave signal due to AMR in the LSMO layer, and (2) spin pumping with detecting the spin current by means of ISHE in the NM. The corresponding contributions will be denoted by $U^{AMR}$ and $U^{SP}$. As seen from Fig.2, the $U(H_0)$ signals contain both symmetric and antisymmetric components. It is known [13] that the shape of $U^{SP}(H_0)$ is always symmetric and coincides with the FMR absorption line. This is not sufficient, however, for its separation from $U^{AMR}$, since the latter may contain both symmetric and antisymmetric components, their ratio being dependent on the phase shift between the microwave field and microwave current [9]. In the geometry used in this work (Fig.1), the microwave current flowing along U is related to some perturbations (see above, Section 2), so its phase depends on particular sample and hardly can be determined *ab* *initio*. As emphasized in Ref. [9], the most reliable way to separate the $U^{AMR}$ and $U^{SP}$ signals is to analyze the angular dependence of the effect when the field $\mathbf{H_0}$ is rotated in the film plane. Note that the antisymmetric signal caused by the anomalous Hall effect [1] is hardly observable because the out-of-plane magnetization component is small due to the strong demagnetization field [9].

The dependences of the magnitude of the symmetric component $U^s$ on the angle $\alpha$ between $\mathbf{H_0}$ and the direction of U measuring for the two samples are presented in Figs. 2(a, b). As seen in the Figure, these data, as well as those for LSMO/Pt (not shown), can be successively approximated by the following expression:

$$U^s(\alpha) = U_0^{AMR} \sin 2(\alpha - \beta) \cdot \sin \alpha + U_0^{SP} \sin^3 \alpha \qquad (1)$$

Two terms in Eq.(1) are in accordance with the theoretically approved formulae for the AMR and SP effects [9] and serve as a base for separating them (note that in our case $H_0 \gg H_u$, so the equilibrium magnetization is directed practically along $\mathbf{H_0}$). The common factor $\sin\alpha$ in Eq. (1) works in any case, since the FMR precession is excited only by the transverse component of $\mathbf{h}$. The curves corresponding to each mechanism are also shown in Fig.2. Maximum voltages for the both effects, as well as the fitting angles $\beta$ for various samples are listed in Table 1. It should be noted that $\beta$ is zero for LSMO/Au and LSMO/Pt, but equals to 40 deg. for LSMO/SRO. Evidently, this is related to the fact that the easy axis coincides with the azimuthal origin ($\alpha=0$) in the first two samples, but differ from it just by 40 deg. in the latter one. Thus, our data point to specific AMR mechanism where the effect depends on the direction of easy axis rather than the direction of electric current.

This conclusion was supported by additional measurements performed with the LSMO film similar to that used in the LSMO/Pt sample but without the NM layer. The $U^{AMR}_{ij}$ signals were detected using different pairs (i,j) of potential contacts situated at the sample corners, see Fig. 3(b). In this case, the registered signals correspond to different directions of the microwave currents contributing to the measured voltage, whereas the azimuthal angle $\alpha$ is counted from the same origin coinciding with the easy axis. It was found that the $U^{AMR}_{12}$, $U^{AMR}_{23}$ and $U^{AMR}_{34}$ signals differed in their shape thus pointing to different phases of microwave currents. At the same time, the dependencies of the signal magnitudes on $\alpha$ [Fig. 3(a)] have the same functional form, being well described by the first term in Eq. (1) at $\beta=0$.

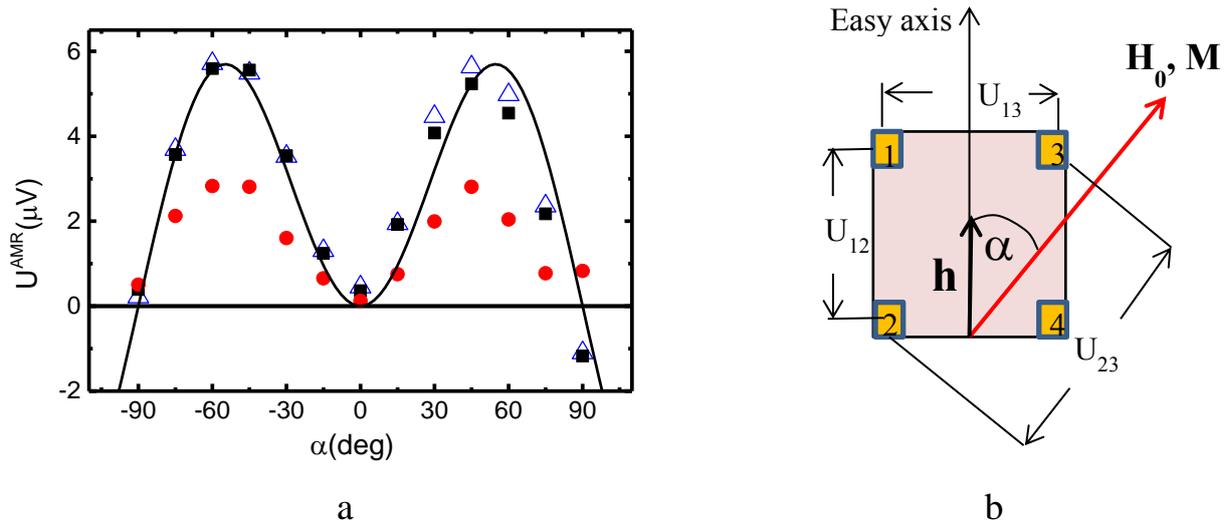

**Fig. 3.** (a) Angular dependences of the potential differences $U_{ij}$ measured between electrical contacts 1,2 (open triangles), 1,3 (filled squares) and 2,3 (circles) in the LSMO film under FMR pumping. The solid curve corresponds to the first term in Eq. (1) at $\beta = 0$. (b) The sketch of the sample.

Temperature dependences of the symmetric components $U^s$ for LSMO/Pt and LSMO/SRO are presented in Fig. 4. The data were taken at different orientations of $H_0$ ($\alpha=\beta$ and $\alpha=15^0$), thus allowing separation of the $U^{AMR}(T)$ and $U^{SP}(T)$ functions. It is seen that both effects decrease steeply when approaching $T_C$.

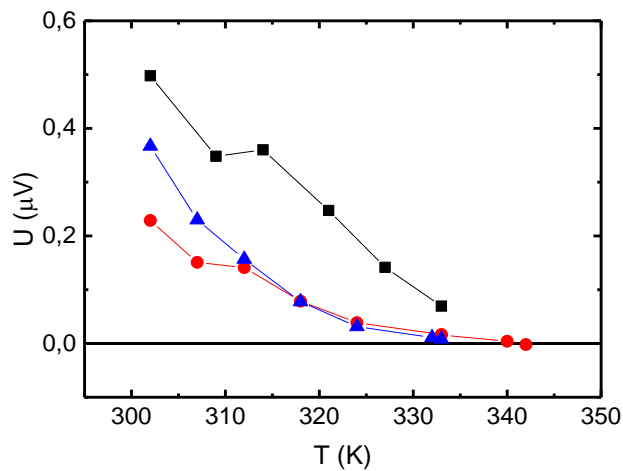

**Fig. 4.** Temperature dependence of the dc voltage caused by spin pumping (LSMO/Pt, filled squares; LSMO/SRO, filled circles) and by AMR effect (LSMO/SRO, triangles). The lines connect the experimental points.

Finally, we present the results of some supplementary measurements to be useful in discussion on the physical mechanism of the resonance e.m.f. in manganites.

The anisotropic magnetoresistance was measured directly in the sample with β = 40⁰. The measuring dc current was directed along the sample edge making an angle β with the easy axis. The dependence of the resistance R on the in-plane **H₀** direction at $H_0 = const. = 1700$ Oe and T=295 K is shown in Fig. 5(a).

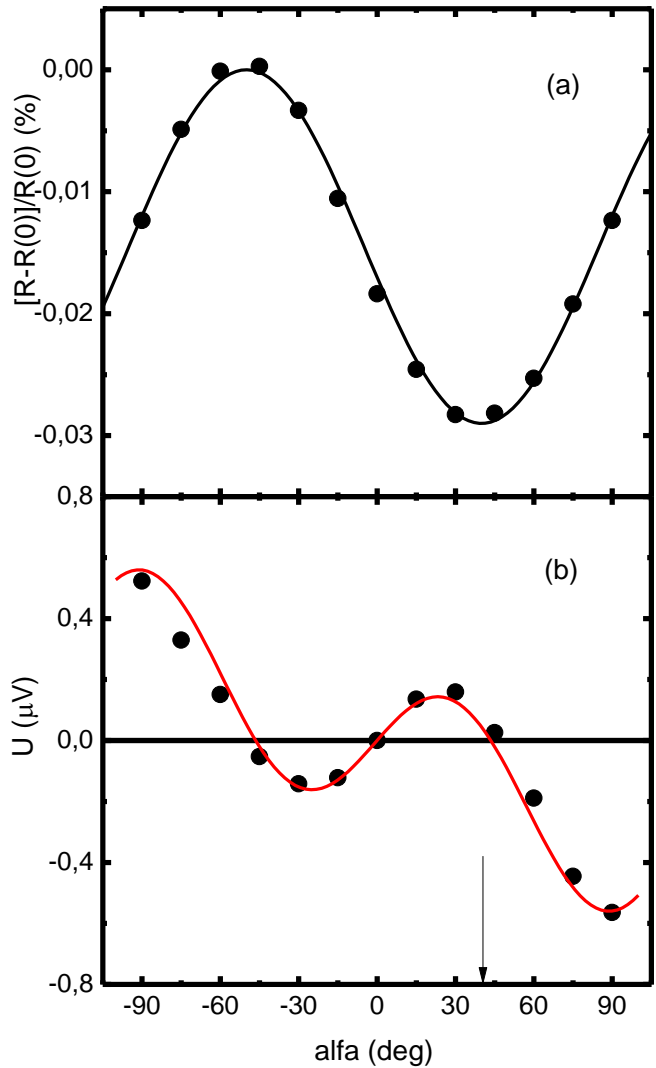

**Fig. 5.** (a) Change in the film resistance measured along the sample edge (α=0) as a function of **H₀** direction in the film plane. The arrow indicates easy axis. Solid curve corresponds to Eq. (2). (b) Angular dependence of $U^{AMR}$ in the same sample. The curve is fitted to the first term in Eq.(1) with β=40⁰.

For comparison, the angular dependence of $U^{AMR}$ is presented in Fig.5(b). As seen from the Fig.5(a), the AMR data can be well described by the expression

$$\frac{R(\alpha)}{R(0)} = 1 + \varepsilon_{AMR}\cos^2(\alpha - \beta) \qquad (2)$$

at $\varepsilon_{AMR} = (- 2{,}9 \pm 0{,}1) \cdot 10^{-4}$. Obviously, this result is consistent with the foregoing treatment of Eq. (1).

Further, the differential CMR factor,

$$r_{CMR} = \frac{dR}{dH} \qquad (3)$$

was measured in the same sample as a function on temperature, see Fig. 6. Note that the R(H) function was found to be approximately linear in the field range used in present experiments; the data shown in Fig. 6 were taken around $H_0 = 2$ kOe.

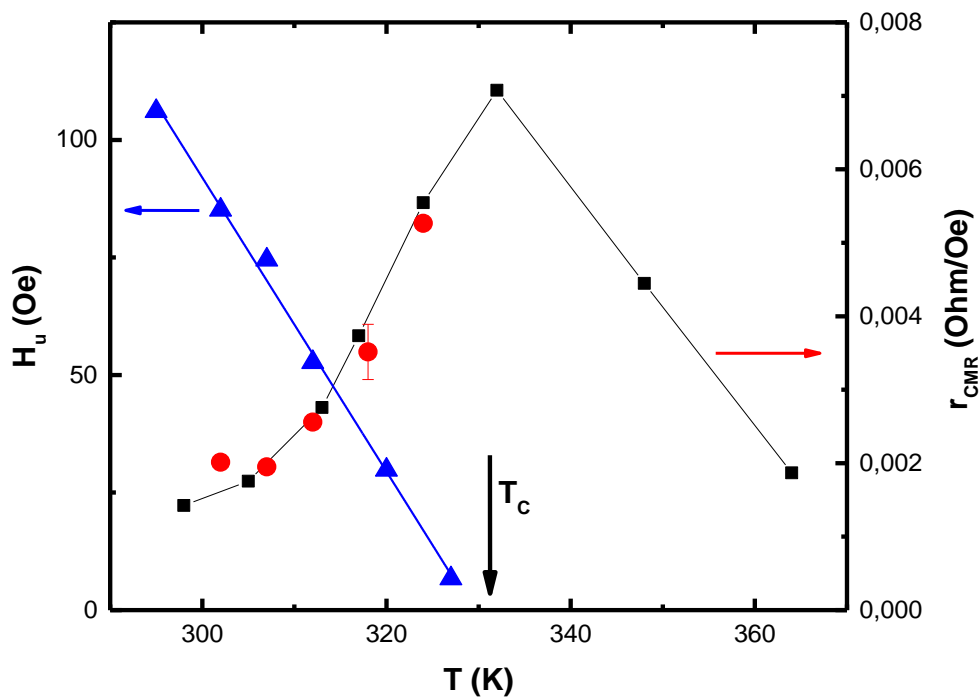

**Fig. 6.** The uniaxial in-plane anisotropy field $H_u$ (triangles, left scale) and differential CMR factor $r_{CMR}$ (squares, right scale) measured in the LSMO film as a function of temperature. Filled circles are the normalized $r_{CMR}$ values calculated with the use of Eq. (8), see the text. The curves are guides for eyes.

In the same Fig. 6, the temperature dependence is plotted of the uniaxial in-plane anisotropy field $H_u$. The $H_u$ values were determined from the angular variation of the FMR line position, see Ref. [28]. As seen from the plot, the anisotropy field falls approximately linearly upon heating and tends to zero near $T_C$. Temperature dependences of electrical resistance R(T), equilibrium magnetization M(T), and FMR line width ΔH(T) were obtained as well, but not shown here for the sake of brevity.

## 4. Discussion

Consider firstly physical mechanism of the resonance e.m.f. caused by anisotropic magnetoresistance in LSMO films. Let us return to the experimental scheme shown in Fig.1. In the simple model [8], the description is based on the Ohm law,

$$U(t) = I(t) \cdot R(t) \qquad (4)$$

where $I(t) = I_0 \sin(\omega t)$ and R(t) are, respectively, the microwave current and electric resistance in the direction of the potential difference to be measured (z-axis in Fig. 1). Resonance microwave pumping leads to precession of magnetization; as a result, the angle α(t) between z-axis and **M**(t) is modulated at the FMR frequency. Note that, owing to demagnetization in a thin film, the precession cone is flattened, so, in the first approximation, one has

$$\alpha(t) = \alpha + \delta \cdot \sin(\omega t + \varphi) \qquad (5)$$

where δ is the cone angle in the film plane and φ is the phase shift relative to the microwave current. At this point, the anisotropic magnetoresistance comes into play. Owing to AMR, the resistance R depends on the magnetization orientation, and hence on the angle α(t). This leads to corresponding modulation of the resistance R(t) and to appearance of the dc component $U^{AMR}$ = <U(t)>. This is the resonance e.m.f. effect under study.

Naturally, the $U^{AMR}$ magnitude depends on α. In the traditional approach suitable for standard elementary ferromagnetic metals and alloys, the anisotropic part of the magnetoresistance is proportional to $\cos^2\alpha$ [3]. In this case, the angular dependence of $U^{AMR}$ would be represented by the first term in Eq. (1) with β=0. However, our experimental data evidence for another behavior. In fact, the effect is not always determined by α, but rather by the deflection of the field **H**$_0$ from easy axis. As a result, the angle α is substituted for (α-β) in Eqs. (1) and (2).

At the first glance, this is consistent with the modern consideration [25, 31] which predicts that the AMR effect in manganites depends on the field direction relative to the [100] crystallographic axis. This approach, however, is restricted by low temperatures (T<<T$_C$) and does not take into account the in-plane magnetic anisotropy; thus, it is hardly applicable to our data.

We propose another interpretation taking into account the colossal magnetoresistance effect (CMR) which is a characteristic property of manganites, especially in the temperature range not far from $T_C$. The CMR effect manifests itself in a decrease in resistivity at $H_0$ increasing; this is commonly related to an increase in spin polarization favoring electron jumps between $Mn^{3+}$ and $Mn^{4+}$ ions [32, 19]. Evidently, such mechanism can be efficient only at high enough temperatures (not too far from $T_C$), when the maximum value of saturated magnetization is not achieved because of thermal fluctuations. Usually CMR raises with temperature in ferromagnetic phase, reaching its maximum near $T_C$, see Fig. 5. It is reasonable to suppose that the magnetization magnitude (the length of the **M** vector) depends not only on the external field **H₀**, but also on internal fields related to demagnetization factor and magnetic anisotropy. All these contributions are included in an effective field **H_e** which can be determined using standard methods with the use of free energy and equilibrium conditions [33]. For the experiments described in the present work, **H_e** is directed practically along **H₀** and equals to

$$H_e = H_0 + H_u \cos^2 \alpha_e \tag{6}$$

where $\alpha_e = (\alpha - \beta)$ is the angle making by **H₀** relative to the easy axis. Since CMR reveals in resistance decreasing at field increasing, Eq. (6) predicts that the minimal value of R should be reached at **H₀** directed along easy axis ([010] in our samples). Recently such a result was obtained experimentally in very thin LSMO films with in-plane uniaxial magnetic anisotropy [34].

To check this model quantitatively, let us compare the decrease in resistivity $\Delta R_{CMR}$ when $H_0$ is increased just by the value of $H_u$ with the resistivity change $\Delta R_{AMR}$ due to **H₀** rotating by $90^0$ from the hard to easy axis. The data obtained are as follows:

$$\Delta R_{CMR} = r_{CMR} H_u = -(0.083 \pm 0.005) Ohm$$

$$\tag{7}$$

$$\Delta R_{AMR} = \varepsilon_{AMR} R = -(0.071 \pm 0.008) Ohm$$

The results coincide within the experimental error, evidencing for validity of the model suggested.

Discuss now the temperature dependence of the resonance e.m.f. effect, Fig.4. According to our model, the voltage $U^{AMR}$ should be represented by an expression analogous to that used in Refs. [8, 9], but with $\Delta R_{CMR}$ substituted for $\Delta R_{AMR}$, see Eq.(7). As a result, we have:

$$U^{AMR}(T) = A \frac{r_{CMR}(T) \cdot H_u(T)}{R(T) \cdot \Delta H(T)} \qquad (8)$$

Here the temperature independent factor A includes microwave field parameters, angular dependence and so on. The inverse proportionality of the microwave current to the resistance R(T) is taken into account.

To check the validity of Eq.(8), the measured values of $U^{AMR}$ (Fig. 4) were multiplied by the expression $R(T)\Delta H(T)/H_u(T)$, the corresponding data being obtained in independent experiments. The results (scaled relative to the ordinate axis) are plotted together with the experimental graph of $r_{CMR}(T)$, Fig. 6. It is seen that the calculated and experimental temperature dependences of $r_{CMR}$ coincide within the experimental error.

Note that substitution of **H$_e$** for **H$_0$** in the AMR interpretation seems to be quite natural and can be compared with similar procedure in the Landau-Lifshits equation [33]. More serious grounds are needed, however, when considering the resonance e.m.f. effect. Indeed, in terms of the suggested approach, resonance oscillations of $\alpha_e$ lead to corresponding modulation of the length of the **M** vector which, in its turn, causes oscillations of resistance R(t) in Eq. (4) owing to the CMR mechanism. Such a model supposes an adiabatic mode of the resonance precession, in the sense that equilibrium values of |**M**| and R are established at every instant of time. This means that the rate of the "absolute" longitudinal spin relaxation along **M** is much more than the precession frequency,

$$\tau_{1abs}^{-1} \gg \omega \sim 6 \cdot 10^{10} s^{-1} \qquad (9)$$

According to Eq. (9), $\tau_{1abs}^{-1}$ exceeds the "ordinary" relaxation rate related to rotating of **M** toward its equilibrium orientation. At the same time, Eq. (9) is consistent with the theory [35, 36] based on the Landau-Lifshits-Bloch equation with account made for thermal fluctuations. As shown in Ref. [36], the rate $\tau_{1abs}^{-1}$ in ferromagnetic phase is determined by exchange interaction and amounts to about $10^{13}$ s$^{-1}$ at the whole temperature range except the close vicinity of $T_C$, where a transition occurs to paramagnetic behavior. Note that this theory was successfully applied to interpretation of another spin-charge effect, namely, resonance magnetoresistance in LSMO films [5, 6].

Passing to discussion on the spin pumping, it should be noted that the magnitude of the effect in the LSMO/NM structures studied in the present work was found to be rather low: the values of $U^{SP}$ at room temperature amount to fractions of µV and decrease steeply when approaching $T_C$. According to the theory [15, 16], the maximum voltage is determined by the expression

$$U_{max}^{SP} = - \frac{e\theta_{SH}}{2\pi(\sigma_{NM}t_{NM}+\sigma_{FM}t_{FM})} \lambda_{SD}\tanh(\frac{t_{NM}}{2\lambda_{SD}})g_{\uparrow\downarrow}\omega Lp(\frac{h}{\Delta H})^2 \qquad (10)$$

where $e$ is the electron charge; $\theta_{SH}$ is the spin Hall angle in NM; $\sigma_{NM}$ ($\sigma_{FM}$) and $t_{NM}$ ($t_{FM}$) are the conductivities and thicknesses of the NM (FM) layers, respectively; $g_{\uparrow\downarrow}$ is the real part of the spin mixing conductance characterizing the FM/NM interface; $\lambda_{SD}$ is the spin diffusion length in NM; $L$ is the sample length along U; and $p \sim 1$ is a factor accounting for ellipticity of spin precession in the film. From Eq. (10), accounting for our experimental data and the metal parameters cited in literature (see, for example, Ref. [24]), the $g_{\uparrow\downarrow}$ value can be estimated as being one or two orders of magnitude less than typical quantities reported for various FI/NM and FM/NM structures [15, 16, 25]. Low $U^{SP}$ voltage obtained in our study might be related to specific features of the LSMO zone diagram, as well as to low spin polarization caused by high enough temperature (see the temperature dependence of $U^{SP}$ in Fig. 4). It seems, however, that the most probable cause of the SP suppression may be defects at the LSMO/NM interface, such as formation of a "dead" layer at the manganite surface. Further studies are planned.

In conclusion, two resonance spin-charge effects, the electromotive force caused by anisotropic magnetoresistance and spin pumping (pure spin current) induced by ferromagnetic resonance excitation have been observed and studied in ferromagnetic manganite epitaxial thin films and manganite/metal bilayers. Temperature dependencies of the effects were traced up to Curie point, where both effects vanished. The two phenomena were separated from each other using variation of the measured voltage as a function on the magnetic field direction. The dc voltage corresponding to the spin pumping was found to be considerably lower than that reported previously for other FM/NM and FI/NM structures. This may be attributed to some peculiarities of the FM/NM interface. Further study of this problem is desirable.

It was also found that the physical mechanism of e.m.f. caused by anisotropic magnetoresistance in the manganite films, as well as the AMR itself, differ considerably from similar effects in traditional ferromagnetic metals. In the manganite films with in-plane magnetic anisotropy, the main role is played by colossal magnetoresistance which provides resistivity variations depending on the angle between **M** and easy axis.

Acknowledgments

The authors are grateful to A.M. Petrzhik, K.I. Constantinian, and Yu.V. Kislinski for their help in experiment. The work was supported by the RFBR grants 14-02-00165, 14-07-93105, and 14-07-00258.

Captions for Figures

**Fig. 1.** The sketch of experimental geometry. The easy axis directions are shown at the left panel for two types of samples (see the text).

**Fig. 2.** Magnitude of symmetric part $U^s$ of the resonance dc voltage as a function of the angle $\alpha$ between the field direction and the sample edge for (a) LSMO/Au and (b) LSMO/SRO. Filled circles are experimental data. Solid curves are the best fits to Eq. (1) with $\beta=0$ (a) and $\beta=40^0$ (b). Dotted and dashed curves correspond to the AMR and SP effects, respectively. Insets: typical $U(H_0)$ signals recorded upon FMR pumping.

**Fig. 3.** (a) Angular dependences of the potential differences $U_{ij}$ measured between electrical contacts 1,2 (open triangles), 1,3 (filled squares) and 2,3 (circles) in the LSMO film under FMR pumping. The solid curve corresponds to the first term in Eq. (1) at $\beta = 0$. (b) The sketch of the sample.

**Fig. 4.** Temperature dependence of the dc voltage caused by spin pumping (LSMO/Pt, filled squares; LSMO/SRO, filled circles) and by AMR effect (LSMO/SRO, triangles). The lines connect the experimental points.

**Fig. 5.** (a) Change in the film resistance measured along the sample edge ($\alpha=0$) as a function of $\mathbf{H_0}$ direction in the film plane. The arrow indicates easy axis. Solid curve corresponds to Eq. (2). (b) Angular dependence of $U^{AMR}$ in the same sample. The curve is fitted to the first term in Eq.(1) with $\beta=40^0$.

**Fig. 6.** The uniaxial in-plane anisotropy field $H_u$ (triangles, left scale) and differential CMR factor $r_{CMR}$ (squares, right scale) measured in the LSMO film as a function of temperature. Filled circles are the normalized $r_{CMR}$ values calculated with the use of Eq. (8), see the text. The curves are guides for eyes.

**Table 1.** Main parameters and experimental data obtained with the bilayer samples under study. $U^{AMR}_{max}$ and $U^{SP}_{max}$ are the peak absolute values of dc signals caused by the anisotropic magnetoresistance and spin pumping effects, respectively. The data shown in six right columns correspond to the room temperature.

| Structure | $t_{FM}/t_{NM}$ (nm/nm) | $T_C$ (K) | M (Oe) | $H_u$ (Oe) | $\Delta H$ (Oe) | $\beta$ (deg.) | $U^{AMR}_{max}$ ($\mu$V) | $U^{SP}_{max}$ ($\mu$V) |
|---|---|---|---|---|---|---|---|---|
| LSMO/Au | 65/15 | 347 | 285 | 85 | 28 | 0 | 0.13 | 0.34 |
| LSMO/Pt | 80/10 | 347 | 302 | 139 | 20 | 0 | 0.2 | 0.5 |
| LSMO/SRO | 60/10 | 331 | 271 | 101 | 80 | 40 | 1.2 | 0.55 |